\documentclass[10pt,aps,twocolumn,superscriptaddress]{revtex4-1}

\usepackage{amsmath}
\usepackage{textcomp}
\usepackage{amssymb} 
\usepackage{graphicx}
\usepackage[usenames]{color}

\makeatletter
\newcommand*{\rom}[1]{\expandafter\@slowromancap\romannumeral #1@} %roman numbers
\makeatother

\begin{document}

\title{Stress-stabilized sub-isostatic fiber networks in a rope-like limit}

\author{Sadjad Arzash}
\affiliation{Department of Chemical \& Biomolecular Engineering, Rice University, Houston, TX 77005}
\affiliation{Center for Theoretical Biological Physics, Rice University, Houston, TX 77030}
\author{Jordan L.\ Shivers}
\affiliation{Department of Chemical \& Biomolecular Engineering, Rice University, Houston, TX 77005}
\affiliation{Center for Theoretical Biological Physics, Rice University, Houston, TX 77030}
\author{Albert J.\ Licup}
\affiliation{Department of Physics \& Astronomy, Vrije Universiteit, Amsterdam, The Netherlands}
\author{Abhinav Sharma}
\affiliation{Leibniz Institute of Polymer Research Dresden, Dresden, Germany}
\author{Fred C.\ MacKintosh}
\affiliation{Department of Chemical \& Biomolecular Engineering, Rice University, Houston, TX 77005}
\affiliation{Center for Theoretical Biological Physics, Rice University, Houston, TX 77030}
\affiliation{Departments of Chemistry and Physics \& Astronomy, Rice University, Houston, TX 77005, USA}
%\affiliation{Department of Physics \& Astronomy,  Vrije Universiteit, Amsterdam, The Netherlands}

\begin{abstract}
The mechanics of disordered fibrous networks such as those that make up the extracellular matrix are strongly dependent on the local connectivity or coordination number. For biopolymer networks this coordination number is typically between three and four. Such networks are sub-isostatic and linearly unstable to deformation with only central force interactions, but exhibit a mechanical phase transition between floppy and rigid states under strain. Introducing weak bending interactions stabilizes these networks and suppresses the critical signatures of this transition. We show that applying external stress can also stabilize sub-isostatic networks with only tensile central force interactions, i.e., a rope-like potential. Moreover, we find that the linear shear modulus shows a power law scaling with the external normal stress, with a non-mean-field exponent. For networks with finite bending rigidity, we find that the critical stain shifts to lower values under prestress.
\end{abstract}

\maketitle

\section{Introduction}

Networks of biopolymers are ubiquitous in biological systems involved in structural and mechanical stability. Examples include cross-linked cortical actin in the cytoskeleton and branched collagenous networks in the extracellular matrix. The underlying local network geometry and the nature of interactions between the constituent fibers play a key role in determining the stability of these networks. Typically, these networks have average coordination or connectivity between three and four, corresponding to branched or cross-linked geometries, respectively. As shown by Maxwell, the isostatic threshold connectivity for linear stability of an interconnected mechanical structure of simple springs is twice the dimensionality for large number of elements, i.e., $z_c=2d$ \cite{maxwell_l._1864}. Based on this argument, biological networks are intrinsically sub-isostatic. Therefore, if the fibers interact only via central forces such as tension and compression, then these networks are unstable with respect to small deformations. Nevertheless, sub-isostatic networks can be rigidified through various stabilizing effects such as strain \cite{alexander_amorphous_1998}, fiber bending interactions \cite{kroy_force-extension_1996,satcher_theoretical_1996,head_deformation_2003,wilhelm_elasticity_2003,das_effective_2007, broedersz_criticality_2011}, active stresses \cite{mizuno_nonequilibrium_2007,koenderink_active_2009,winer_non-linear_2009,jansen_cells_2013,sheinman_actively_2012}, or thermal fluctuations \cite{plischke_entropic_1998,dennison_fluctuation-stabilized_2013,jaspers_ultra-responsive_2014} giving rise to a stable linear elastic response.

Biopolymer networks also exhibit striking nonlinear elasticity: with barely a 10\% increase in strain, the stiffness increases by almost two orders of magnitude. Such nonlinear mechanics are observed for intracellular cytoskeletal filaments, extracellular fibrin clots, and even whole tissues \cite{mackintosh_elasticity_1995,shah_strain_1997,shadwick_mechanical_1999,gardel_elastic_2004,storm_nonlinear_2005,pritchard_mechanics_2014,arevalo_stress_2015}. The nonlinear mechanics of athermal networks have been described theoretically in terms of a crossover from bending-dominated to a stretching-dominated response \cite{onck_alternative_2005,heussinger_stiff_2006}, normal stresses \cite{licup_stress_2015}, or strain-controlled critical phenomena \cite{sharma_strain-controlled_2016}.
In these theoretical approaches, bending rigidity provides stability in the linear regime. 
It is also known experimentally that polymerization of hydrogels such as collagen and fibrin generally results in prestress \cite{van_oosten_uncoupling_2016}. In fact, some prestress is almost inevitable as crosslinks form between fibers \cite{huisman_monte_2008}.
But, it is still not well understood how such prestress, either externally applied or due to internal constraints, affects network stability and nonlinear mechanics \cite{vos_programming_2017}.
In the case of active stresses, such as by myosin motors in the cytoskeletal or platelet contraction in blood clots, such active prestress can give rise to shear moduli that can exceed the passive shear modulus of the underlying substrate \cite{koenderink_active_2009,winer_non-linear_2009}.\\ \indent Here, in order to investigate the stabilization effect of prestress, we study sub-isostatic rope networks without bending interactions. The elastic response of a rope-like fiber is governed purely by central-force interactions under tension and has vanishing resistance under compression. This represents a minimal model for athermal fibers with zero bending rigidity, for which the Euler buckling threshold vanishes. The external normal stress is applied by either bulk or uniaxial expansion. We show that the linear shear modulus scales as power law with the imposed external normal stress, with a non-mean-field exponent which leads to a divergent susceptibility. This suggests that sub-isostatic rope networks become infinitely susceptible to any stress that invokes fiber stretching modes, including the self-generated normal stresses. We also show that the mechanics of stress-stabilized rope networks can be captured in terms of these normal stresses. Furthermore, by calculating the non-affine fluctuations for both prestressed rope and bend-stabilized networks, we show that prestressing removes criticality from sub-isostatic rope networks and shifts the critical point to lower values for a bend-stabilized network.
\section{Model}
\begin{figure}
	\includegraphics[width=8cm,height=8cm,keepaspectratio]{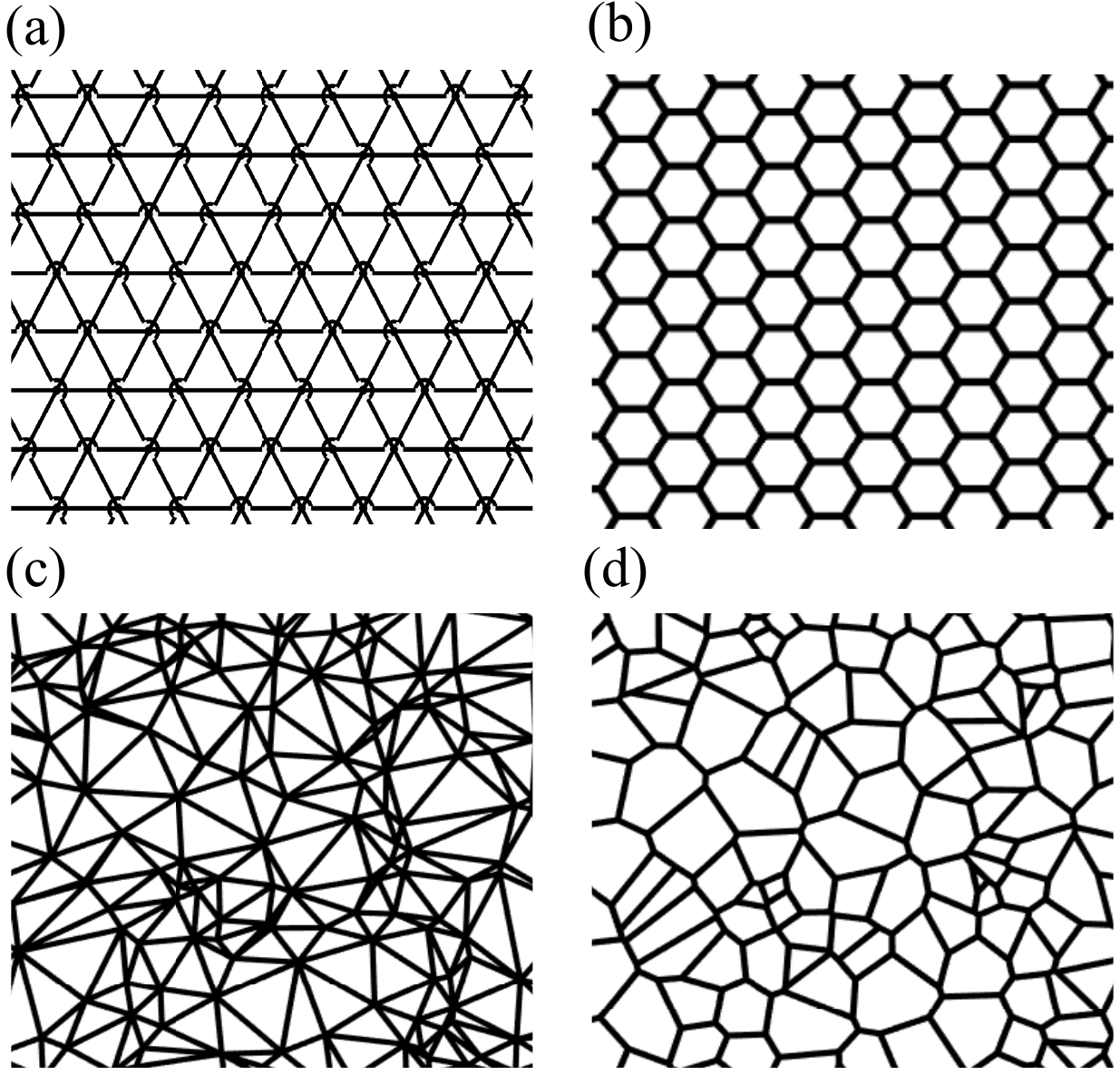}
	\caption{ \label{fig:1} Different geometries used to model fiber networks. (a) A full phantom triangular lattice which has a connectivity of $z=4$. The arcs specify that one of the three crossing fibers has been detached from the crosslink, i.e., it is phantomized. (b) Showing a full hexagonal (honeycomb) lattice which gives a connectivity of $z=3$. (c) A Delaunay network of a random point set. Delanauy network has a non-uniform local connectivity. The average connectivity is 6. (d) A Voronoi network of random points which has an average connectivity of $z=3$. The Voronoi diagram has uniform local connectivity of 3.}
\end{figure}

We use the \textit{phantom} triangular lattice model \cite{broedersz_molecular_2011,broedersz_filament-length-controlled_2012,broedersz_modeling_2014} to study fibrous networks in the rope limit. The networks are generated on a periodic 2D triangular lattice with lattice spacing $l_0=1$ and freely-hinging crosslinks at intersection points. The lattice occupies an area $A=aW^2$, where $W$ is the network size and $a$ is the area of a unit cell. The full triangular network has local connectivity of 6. In real biopolymer networks, the local connectivity can be either 3, corresponding to a branching point, or 4, corresponding to a crosslink between two fibers, so the maximum average connectivity cannot exceed 4.  To satisfy this constraint, we randomly detach one of three fibers at every crosslink on a triangular lattice, reducing the average connectivity from 6 to 4. Moreover, we cut a single bond on each fiber at a random location to remove the unphysical effects of network-spanning fibers. Since there are $3\times W$ fibers on a triangular lattice and the average network connectivity $z$ is calculated as twice number of bonds divided by number of nodes, this fiber cutting step gives a connectivity $z=4-\frac{6}{W}$ which approaches 4 in the limit of large $W$. We then remove random bonds to obtain the desired connectivity. Dangling ends, which have no effect on the mechanical behavior of the network, are removed. Figure \ref{fig:1}a shows a small section of a full phantom triangular model.

In order to compare our results with other geometries, we use three additional 2D network structures: (i) fully branched hexagonal (honeycomb) lattice \cite{rens_nonlinear_2016} (ii) Delaunay triangulation network \cite{delaunay_sur_1934,gurtner_stiffest_2014} and (iii) a Voronoi network \cite{dirichlet_uber_2009,voronoi_nouvelles_1908,heussinger_role_2007}. We note that the notion of \emph{fibers} in these structures is less well-defined than in the context of a triangular lattice. The honeycomb network is easily derived from a full triangular lattice by cutting specific bonds. The Delaunay networks are constructed by placing $N$ random points in a $W\times W$ box and triangulate them in a way that there is no point inside the circumcircle of any triangle (the unique circle passing through the three vertices of the triangle) which maximizes the smallest angle among
all triangulations of the given point set \cite{delaunay_sur_1934}. We use $N=W^{2}$ to obtain an average bond length close to $1$ similar to the triangular lattice model. An interesting aspect of a full Delaunay structure is that it has, by construction, a non-uniform local connectivity, in contrast to a uniform structure of a full triangular lattice. The average connectivity of a full Delaunay network, however, is 6 similar to a triangular lattice. To achieve a sub-isostatic Delaunay network ($z<4$) , we randomly remove bonds with no phantomization. Voronoi networks are derived from Delaunay networks of $N=W^{2}/2$ random points by connecting the centers of the circumcircles which gives an average bond length close to $1$ similar to the honeycomb lattice structure. Like the honeycomb lattice, Voronoi networks have a uniform local connectivity of 3. To remove edge effects, we impose periodic boundary conditions in both directions for all networks and utilize Lees-Edwards boundary conditions \cite{lees_computer_1972} to apply shear strain. We use the network size of $W=100$ for phantom triangular, $W=120$ for honeycomb, and $W=70$ for Delaunay and Voronoi models. In order to obtain sufficient statistics, we use $50$ different realizations for every simulation. Moreover, to remove possible effects of underlying anisotropy, we average the quantities over both the positive and negative strain directions.

\begin{figure*}
	\includegraphics[width=16cm,height=16cm,keepaspectratio]{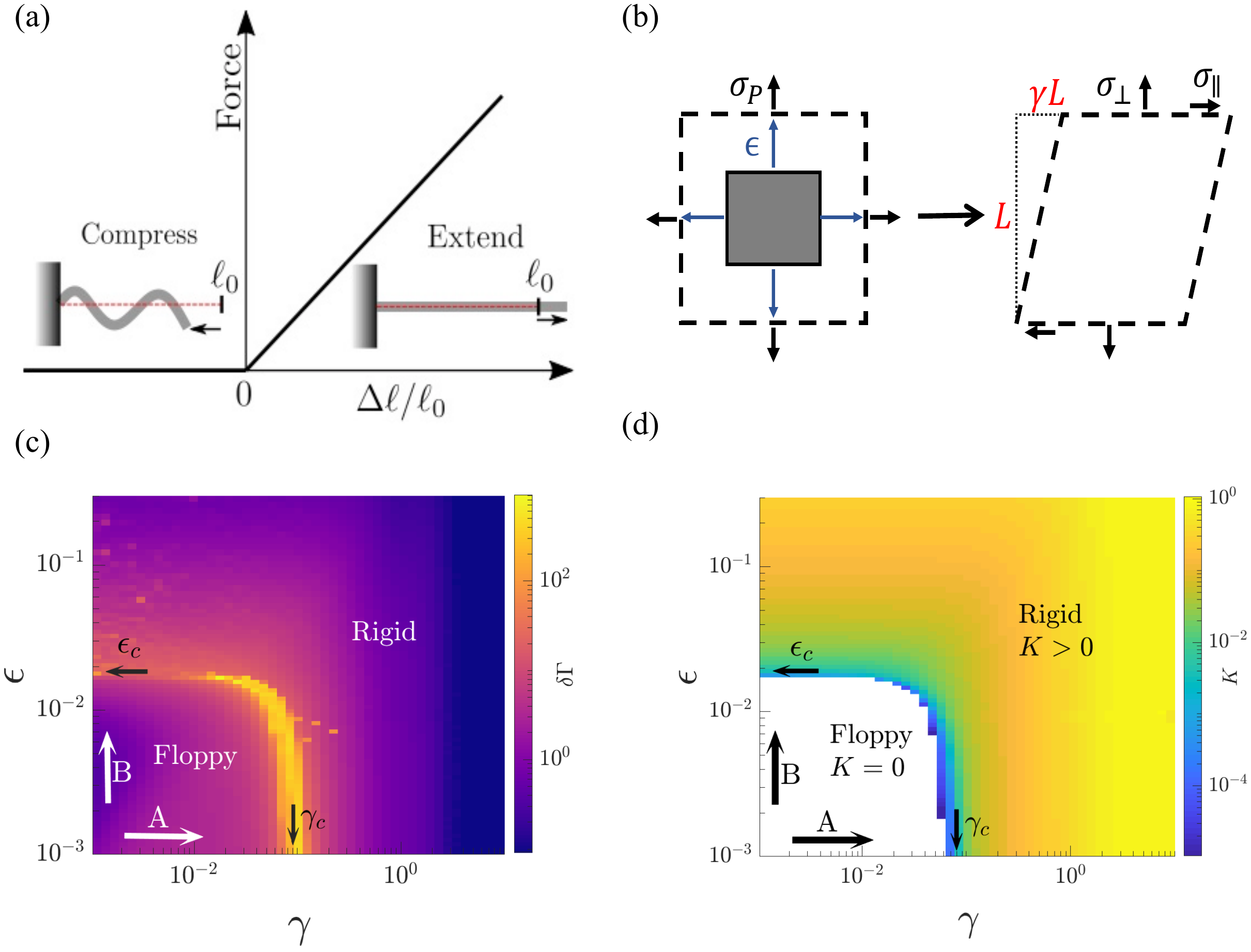}
	\caption{ \label{fig:2} Description of the rope limit and corresponding phase diagram. (a) Showing a schematic force-extension curve for a segment with backbone length $\ell_0$ in the rope limit. The segment behaves as a simple Hookean spring under extenstion and has zero resistance under compressive loads. (b) A schematic of the simulation procedure in the case of prestressing a network by bulk expansion. We first affinely apply a bulk strain $\epsilon$ to the original network that is shown schematically as a black square. The prestress $\sigma_P$ is calculated after finding the minimum energy configuration, allowing for non-affine deformations. To find the shear properties of the network under prestress $\sigma_P$, we affinely shear the expanded network and minimize its elastic energy (see the sketch at right side of the figure). We shear the network in multiple steps, find differential shear modulus $K = \frac{\partial \sigma_\parallel}{\partial \gamma}$ and the linear shear modulus $G = K(\gamma \rightarrow 0)$. (c) The phase diagram of sub-isostatic fiber networks in a rope-like potential. The data is for a phantom triangular network with connectivity of $z=3.2$ that is prestressed by bulk expansion. For small amount of shear strain $\gamma$ or bulk strain $\epsilon$, the network is unstable. However, applying large shear (Arrow $\mathbf{A}$) or extensional strain (Arrow $\mathbf{B}$) removes floppy modes and stabilizes the network. The phase transition from floppy to rigid is captured by showing the differential non-affinity parameter (see Eq.\ 5 in the text), which measures the non-affine fluctuations of the network crosslinks. As expected, the transition curve corresponds to large non-affine fluctuations. The asymmetric behavior of $\delta\Gamma$ under volumetric strain $\epsilon$ and shear strain $\gamma$ is due to the fact that we measure fluctuations under shear strain as can be seen from $\delta\Gamma$ definition. (d) The phase diagram of data in (c) in terms of the differential shear modulus $K$. Floppy networks have no resistance under deformations ($K = 0$) and are rigidified by applying shear or extentional strain larger than a critical value. The phase boundary is the same as the boundary we find in (c) by looking at the fluctuations.}
\end{figure*}

The energy of these networks has two main contributions: stretching of individual bonds and bending between nearest-neighbor bonds (collinear bonds in phantom triangular networks). Therefore, the total energy of the network with stretching stiffness $\mu$ and bending stiffness $\kappa$ is written as
\begin{equation}\label{eq:1}
	\mathcal{H} = \frac{\mu}{2} \sum_{ \langle ij \rangle}^{}\frac{ (l_{ij} - l_{ij,0}) ^2 }{l_{ij,0}} + \frac{\kappa}{2} \sum_{ \langle ijk \rangle}^{} {\frac{ (\theta_{ijk} - \theta_{ijk,0} )^2}{ \frac{1}{2}(l_{ij,0} + l_{jk,0}) } }
\end{equation}
where $l_{ij,0}$ and $l_{ij}$ are the initial and current bond length between crosslinks $i$ and $j$, respectively, and $\theta_{ijk,0}$ and $\theta_{ijk}$ are the initial and current angle between neighboring bonds $ij$ and $jk$, respectively. For networks with finite stiffness, we set $\mu=1$ and vary the dimensionless bending stiffness $\tilde{\kappa} = \frac{\kappa}{\mu l_0^2}$. In the rope limit, however, the total energy depends only on extension of bonds, i.e., we remove bending and also compressive terms from the above energy expression which gives
\begin{equation}\label{eq:2}
\mathcal{H} = \frac{\mu}{2} \sum_{ \langle ij \rangle}^{} \Theta(l_{ij} - l_{ij,0})\frac{ (l_{ij} - l_{ij,0}) ^2 }{l_{ij,0}}
\end{equation}
where $\Theta(x)$ is the Heaviside step function. This is indeed an extreme limit of an asymmetric Hookean spring which has spring constant $\mu$ in the extended state and no resistance under compression. Figure \ref{fig:2}a shows the force-extension curve for a rope segment with insets describing stretching and compression of the segment with original length $l_0$ under extension $\Delta\ell = \ell - \ell_0$. After applying a deformation, we minimize the total energy of the network using FIRE algorithm \cite{bitzek_structural_2006}. The stress components are calculated using the microscopic definition of stresses in a polymeric system as discussed in \cite{Doi_Edwards,shivers_scaling_2018}
\begin{equation} \label{eq:3}
	\sigma_{\alpha \beta} = \frac{1}{2A} \sum_{ \langle ij \rangle}f_{ij,\alpha}r_{ij,\beta}
\end{equation}
where $A$ is the area of the simulation box, $f_{ij,\alpha}$ is the $\alpha$-component of the force exerted on crosslink $i$ by crosslink $j$, and $r_{ij,\beta}$ is the $\beta$-component of the displacement vector connecting crosslinks $i$ and $j$.
In order to investigate the effect of external stresses on a sub-isostatic rope network, we induce finite normal stresses by applying either bulk or uniaxial extension. After applying bulk or uniaxial strain to induce a finite external normal stress, we investigate the shear rheology of the network by applying incremental shear strains to the prestressed network. This procedure is schematically shown in Fig.\ \ref{fig:2}b.

The phase diagram of sub-isostatic networks in the rope limit is shown in Fig. \ref{fig:2}c and Fig. \ref{fig:2}d by looking at non-affine fluctuations and differential shear modulus respectively. In the case of simple shear strain only, the network is floppy below an applied critical strain that is a function of network connectivity and geometry. Beyond this critical point, stretching modes rigidify the network (Arrow \textbf{A} in Fig. \ref{fig:2}c). Although volume-preserving shear deformation of sub-isostatic networks has been extensively studied in prior work, this strain-induced rigidification occurs under \textit{any} type of applied extensional strain; indeed, we see similar phase transition between unstable and stable states under both isotropic expansion and uniaxial extension (Arrow \textbf{B} in Fig. \ref{fig:2}c). This strain-controlled transition occurs due to tension propagation between boundaries that generates a state of self-stress, therefore stabilizes the network \cite{vermeulen_geometry_2017}. To test this in the rope limit, we deform networks by either isotropic expansion or uniaxial extension until the network develops a finite (normal) prestress $\sigma_P$ and apply step-wise shear strains in the direction of arrow \textbf{A} in Fig. \ref{fig:2}c. The isotropic case mimics the uniform active stresses generated by motor proteins or cell contractility on a fibrous network substrate. The second case is motivated by axial expansion or compression experiments of biopolymer gels surrounded by buffer, where the solvent freely flows in or out of the sample thereby preserving the gel boundaries \cite{van_oosten_uncoupling_2016}. Since the uniaxial extension is applied in $y$-direction, throughout this paper, $\sigma_P=\sigma_{yy}$ refers to the normal prestress generated by uniaxial exension prior to a step-wise shear deformation, likewise $\sigma_P =\sigma_{B}= \frac{1}{2} (\sigma_{xx} + \sigma_{yy})$ in the case of bulk expansion and $\sigma_{\perp} = \sigma_{yy}$ denotes the generated normal stress during shear deformations which is equal to $\sigma_P$ at $\gamma = 0$.

\section{Results}

\begin{figure*}
	\includegraphics[width=16cm,height=16cm,keepaspectratio]{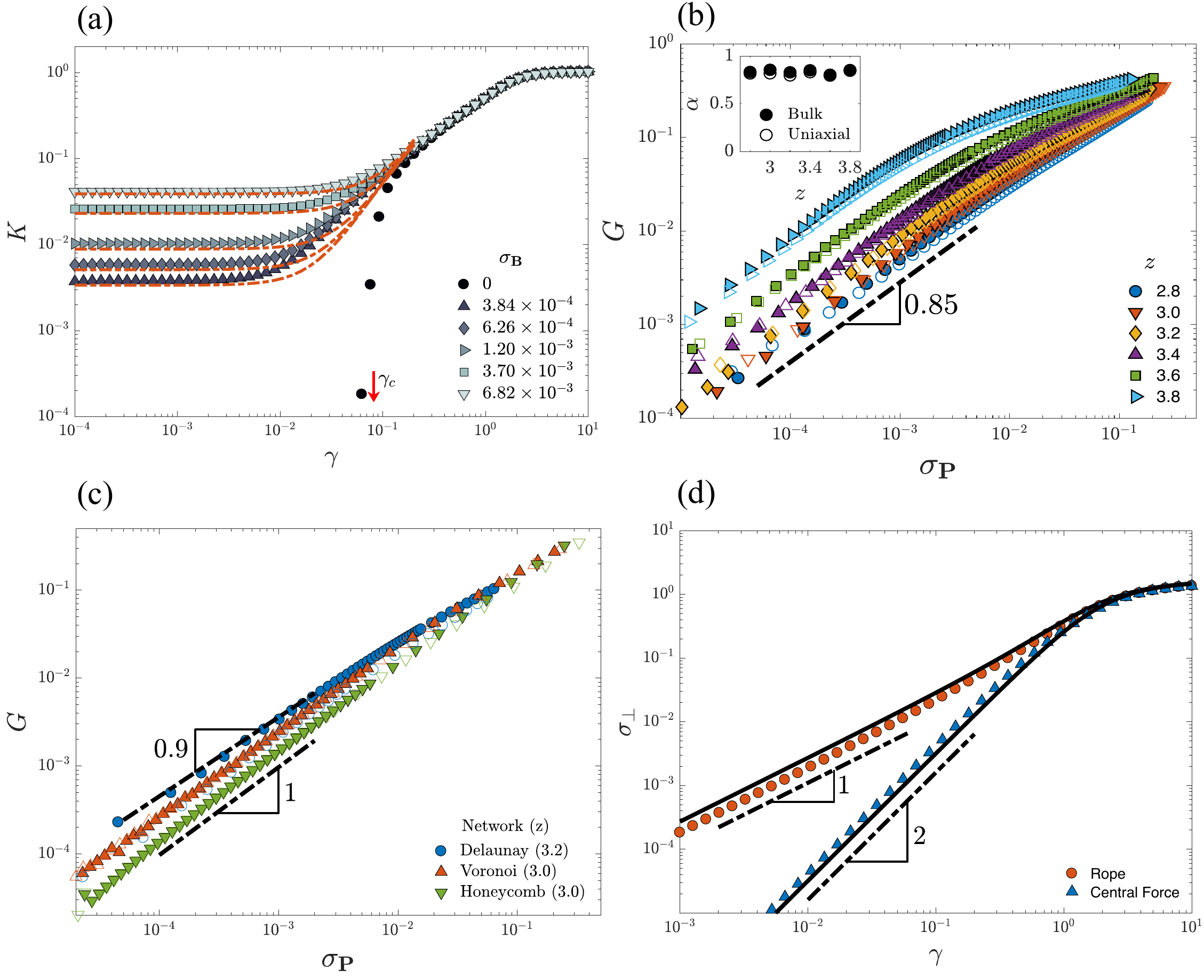}
	\caption{ \label{fig:3} External normal stress stabilizes sub-isostatic fiber networks in the rope limit. (a) Differential shear modulus $K$ versus shear strain $\gamma$ for various amount of external normal stress imposed by bulk expansion on a phantom triangular rope network with average connectivity of $\langle z \rangle \simeq 3.2$ and size $W^2 = 100^2$. In the absence of prestress, the network is floppy below the critical shear strain $\gamma_c$, which is indicated by the red arrow in the figure. However, by applying sufficient bulk expansion to induce finite external normal stress $\sigma_B$, the network becomes stable under shear deformation and exhibits finite $K$. Dashed lines are the results of the stiffening relation (Eq. \ref{eq:4} in the text). 
(b) Linear shear modulus $G$ (obtained as $K$ in the linear regime, with $\gamma\simeq10^{-4}$) versus applied normal stress for varying average connectivity $\langle z\rangle$ of phantom triangular rope networks. The external normal stress $\sigma_P$ is imposed by both bulk (closed symbols) and uniaxial expansion (open symbols). In the small prestress regime, we see a sublinear scaling $G \sim \sigma_P^\alpha$ with $\alpha \simeq 0.85$ which is shown as dashed line in the figure. Inset: showing the scaling exponent $\alpha$ versus network connectivity $z$ which is obtained by fitting a power law to small prestress data $\sigma_P \lesssim 3\times10^{-3}$. This exponent shows no dependence on the network connectivity. Moreover, prestressing the network via bulk or uniaxial extension appears to give the same scaling exponent. (c) Linear shear modulus $G$ obtained by the same procedure as in (b) for different network geometries in the rope limit. The structures with uniform local connectivity, i.e., full honeycomb and Voronoi exhibit an apparent mean-field scaling exponent of $\alpha = 1.0$, in contrast the disordered Delaunay and phantom triangular networks exhibit a non-mean field behavior.
(d) Normal stress $\sigma_{\perp}=\sigma_{yy}$ versus shear strain for a super-isostatic Delaunay network with $z=6.0$ and no applied external normal prestress. As expected, for a network with pure central force interactions the normal stress is quadratic in shear strain, however, for a rope network this relation is linear. The solid lines are calculated using the pure affine isotropic network model as discussed in \cite{storm_nonlinear_2005} using either rope or Hookean spring force-extension relations.}
\end{figure*}

Figure \ref{fig:3}a shows the differential shear modulus $K = \frac{d\sigma_{\parallel}}{d\gamma}$ where $\sigma_{\parallel}$ is the shear stress, versus shear strain $\gamma$ for a sub-isostatic rope network ($z=3.2$) for different amounts of external normal stress $\sigma_B$ caused by bulk expansion. In the absence of external stress ($\sigma_B=0$), the sub-isostatic rope network has no resistance under small shear deformation. Applying sufficient bulk expansion to induce finite $\sigma_B$ (crossing the phase boundary along the $\epsilon$ axis in Fig. \ref{fig:2}c) stabilizes the network, resulting in a finite shear modulus in the linear regime, and further increasing $\sigma_B$ leads to an increase in $K$ in the linear regime. Similar behavior is observed in sub-isostatic fiber networks with finite bending stiffness $\kappa$ when $\kappa$ is increased \cite{sharma_strain-controlled_2016,licup_stress_2015,licup_elastic_2016,sharma_strain-driven_2016} in the absence of applied bulk strain (below the phase boundary in Fig. \ref{fig:2}c). Despite this similarity between stress-stabilized and bend-stabilized networks, the microscopic picture of these two mechanisms is intrinsically different. Bend-stabilized networks resist deformations in the linear regime due to their bending stiffness $\kappa$, whereas stress-stabilized rope networks have already crossed the phase boundary ($\epsilon \geq \epsilon_c$ in Fig. \ref{fig:2}c) and thus show resistance under small shear strains because of highly stretched segments. We find that in rope networks under either bulk or uniaxial expansion, the linear shear modulus $G=\lim_{\gamma\to0}K$ increases with the external normal stress $\sigma_P$ as a power law, $G \sim \sigma_P^\alpha$, with a non-mean-field exponent $\alpha$ (see Fig. \ref{fig:3}b). As expected, stress-stabilized networks with higher average connectivity $z$ show a larger linear shear modulus under equivalent $\sigma_P$. As shown in the inset of Fig.\ \ref{fig:3}b, we find a very weak dependence of this scaling exponent on the network average connectivity $z$. Moreover, this scaling exponent appears to be independent of the prestressing method we used, i.e., bulk or uniaxial extension. As shown in Fig. \ref{fig:3}b, the linear shear modulus of different network connectivity under large prestress deviates substantially from the power law scaling and has a converging trend. This is due to the fact that network segments are massively stretched under large expansion steps and hence the linear shear modulus is primarily governed by this stretching load rather than connectivity or density of networks. The significance of a sublinear scaling suggests the role of prestress in stabilizing a rope network. We define a susceptibility to the applied prestress, $\chi_P \propto \frac{dG}{d\sigma_P} \sim \sigma_P^{\alpha -1}$, which diverges as $\sigma_P \rightarrow 0$ for $\alpha<1$. In the small strain regime $\gamma < \gamma_c$, we find that the excess normal stress $|\sigma_{\perp} - \sigma_P|$ generated under shear remains negligible ($\lesssim 10^{-5}$) with fixed axial strain $\epsilon$. This is in contrast with fiber networks stabilized by finite bending modulus $\kappa$ or super-isostatic networks with only central force interactions, where in the linear regime, the normal stress in the shear deformation is quadratic , i.e., $\sigma_{\perp} \sim \gamma^2$ \cite{janmey_negative_2007,conti_cross-linked_2009,licup_elastic_2016}. Interestingly, normal stresses in the shear deformation of a super-isostatic rope network, which is stable due to the large number of constraints, are proportional to shear strain in the linear regime, i.e., $\sigma_{\perp} \sim \gamma$ (see Fig. \ref{fig:3}d). This is due to the fact that the symmetry of the potential is broken in the rope limit. From these observations, we propose the following nonlinear stiffening relation in the stress-stabilized sub-isostatic rope networks \cite{licup_stress_2015}
\begin{equation}\label{eq:4}
	K \sim \chi_P \sigma_{\perp}, \; \gamma \leq \gamma_c
\end{equation}
in which $\gamma_c$ refers to the critical strain of a rope network in absence of any prestress.
Since in the linear regime $\sigma_{\perp} \sim \sigma_P$, the above relation results in the sublinear scaling relation $G \sim \sigma_P^\alpha$. The dashed lines in Fig.\ \ref{fig:3}a show predictions of the above stiffening relation.

The scaling relation $G \sim \sigma_P^\alpha$ with a non-mean-field exponent $\alpha<1.0$ signifies the influence of the non-mean-field characteristics of sub-isostatic phantom triangular networks, i.e their disordered structure and inhomogeneous local connectivities. Networks with more ordered geometries are expected to show mean-field behavior. To investigate this, we simulate full honeycomb, full Voronoi, and diluted Delaunay networks in the rope limit. As expected, networks with uniform local connectivity of 3, i.e., full honeycomb and Voronoi, exhibit mean-field $\alpha=1$ in contrast to the diluted Delaunay network, with inhomogeneous local connectivity, which yields a non-mean-field $\alpha \cong 0.9$ as the diluted phantomized triangular networks (see Fig. \ref{fig:3}c).

\begin{figure}
	\includegraphics[width=11cm,height=11cm,keepaspectratio]{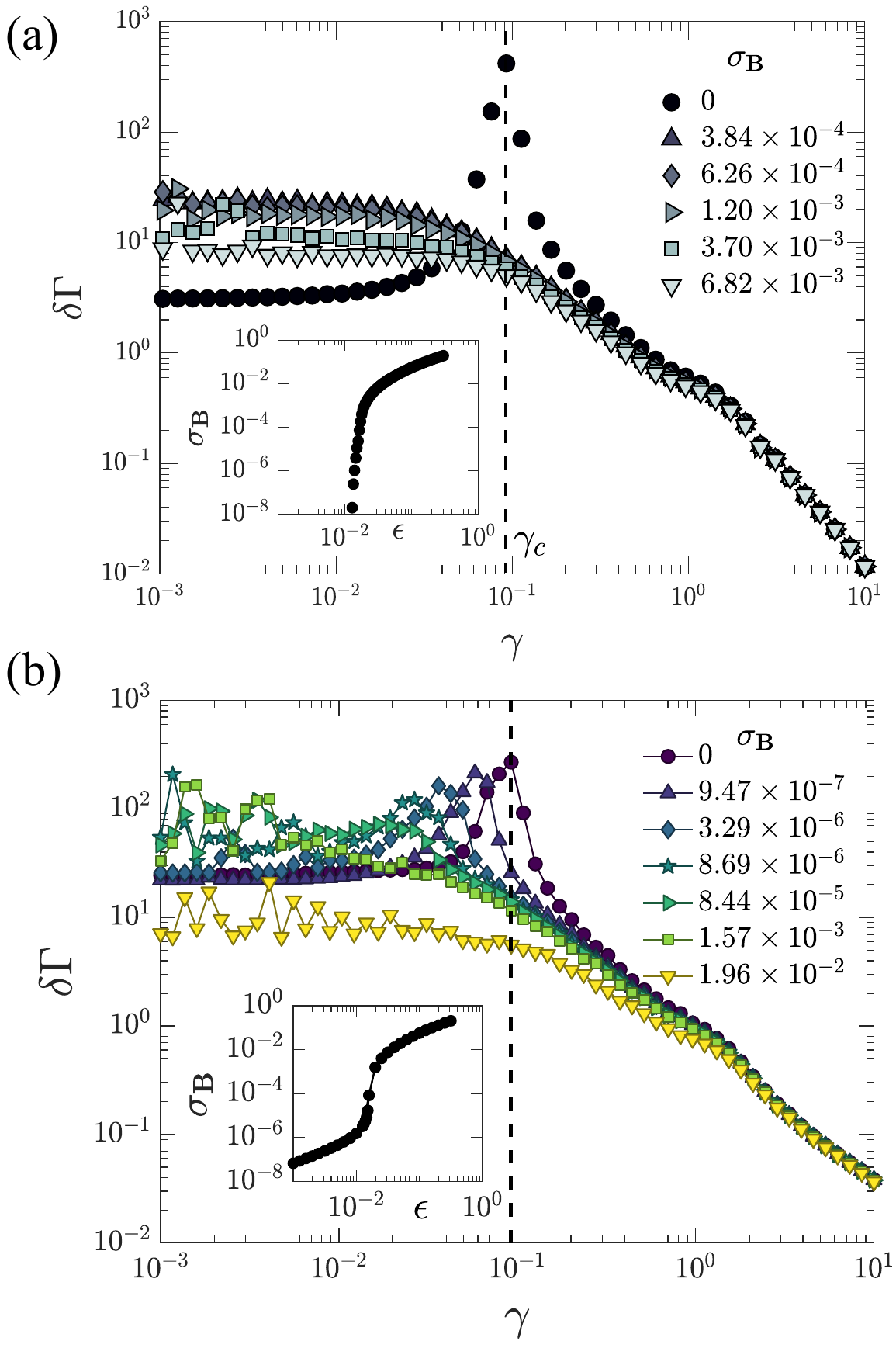}
	\caption{ \label{fig:4} Non-affine fluctuations of sub-isostatic networks both in the rope limit and in presence of bending interactions. (a) Differential non-affinity calculated for phantom triangular networks with connectivity $z=3.2$ in the rope limit for different external stress which is applied by isotropic expansion. The suppression of non-affine deformations due to the external normal stresses is clearly observed. Inset: showing the external stress versus volumetric strain $\epsilon$. Sub-isostatic networks with rope-like potential are unstable under small strains $\epsilon < \epsilon_c$. (b) Showing differential non-affinity for the same network in (a) in presence of bending interactions. We used the dimensionless bending stiffness of $\tilde{\kappa} = 10^{-6}$. Small applied external stress $\sigma_B$ shifts the critical strain $\gamma_c$ to lower values. Applying sufficient extension to drive these networks above the critical extension, like in stress-stabilized rope networks, removes the peak in $\delta\Gamma$. Inset: showing the external stress versus volumetric strain $\epsilon$ for networks with bending interactions. Due to bending interactions, the networks are stable and their behavior can be captured under any small amount of applied prestress. Similar to rope networks, the non-affine fluctuations in bend-stabilized networks are suppressed under large volumetric strain $\epsilon > \epsilon_c$.}
\end{figure}

The transition between floppy (bending-dominated) and rigid (stretching-dominated) states of sub-isostatic fiber networks has been studied in the absence (presence) of bending interactions \cite{shivers_scaling_2018,sharma_strain-controlled_2016,sheinman_nonlinear_2012,licup_stress_2015,feng_nonlinear_2016,sharma_strain-driven_2016,van_oosten_uncoupling_2016,vahabi_elasticity_2016,vermeulen_geometry_2017,jansen_role_2018,wyart_elasticity_2008,rens_nonlinear_2016,rens_micromechanical_2018,merkel_minimal-length_2018,silverberg_structure-function_2014}. Many of these prior studies have shown that this transition is accompanied by critical signatures, such as the divergence of the non-affine (inhomogeneous) fluctuations in the strain field. Although prior work has discussed possible discontinuity of the modulus at the transition \cite{vermeulen_geometry_2017, merkel_minimal-length_2018}, it is important to note that such a discontinuity does not alter the critical nature of this strain-controlled transition. Since it involves a second derivative of the energy with respect to the control variable strain, the modulus can be thought of as analogous to the heat capacity \cite{shivers_scaling_2018}, which can be discontinuous at a critical point.

Similar to previous studies \cite{sharma_strain-controlled_2016}, we define the differential non-affinity as
\begin{equation}\label{eq:5}
	\delta \Gamma = \frac { \langle \parallel \delta \mathbf{u}^{\text{NA}} \parallel ^2 \rangle}{l^2 \delta\gamma^2}
\end{equation}
where $l$ is the typical bond length of the network and $ \delta \mathbf{u}^{\text{NA}} = \mathbf{u} - \mathbf{u}^{\text{affine}}$ is the differential non-affine displacement of a crosslink caused by applying a small amount of shear strain $\delta\gamma$. We find the average of this quantity over all crosslinks in the network. Like in sub-isostatic spring networks, sub-isostatic rope networks in the absence of any external stresses show a mechanical phase transition under simple shear deformation between floppy and rigid states (Arrow \textbf{A} in Fig. \ref{fig:2}c) coinciding with a peak in the non-affine fluctuations. Without any applied prestress, $\delta\Gamma$ of a sub-isostatic rope network shows a large peak at the critical strain $\gamma_c$ where the stretching energy becomes finite and stabilizes the network (see Fig. \ref{fig:4}a). The critical strain has a strong dependence on the network connectivity $z$ as has been discussed previously \cite{sharma_strain-controlled_2016,licup_elastic_2016,shivers_scaling_2018}. Applying any finite external stresses by either isotropic or uniaxial expansion, however, removes the criticality signatures and therefore $\delta\Gamma$ shows no peak as shown in Fig. \ref{fig:4}a. This is due to the fact that by imposing large expansion prior to shear deformation (Arrow \textbf{B} in Fig. \ref{fig:2}c), we indeed move the network out of floppy state and stabilize it. Moreover, applying small extensional strain $\epsilon<\epsilon_c$ leads to a decrease in $\gamma_c$ with $\gamma_c=0$ for $\epsilon\ge\epsilon_c$. This effect can be captured more easily in a bend-stabilized network in which finite $\sigma_P$ occurs for $\epsilon<\epsilon_c$. As shown in Fig. \ref{fig:4}b, for bend-stabilized sub-isostatic fiber networks, the applied prestress by bulk expansion $\sigma_{B}$ clearly shifts the critical strain to lower values until a point where $\sigma_{B}$ is large enough, i.e., $\epsilon\ge\epsilon_c$ to transform the bending-dominated to stretching-dominated state and removes criticality from the network.

\begin{figure}
	\includegraphics[width=7cm,height=7cm,keepaspectratio]{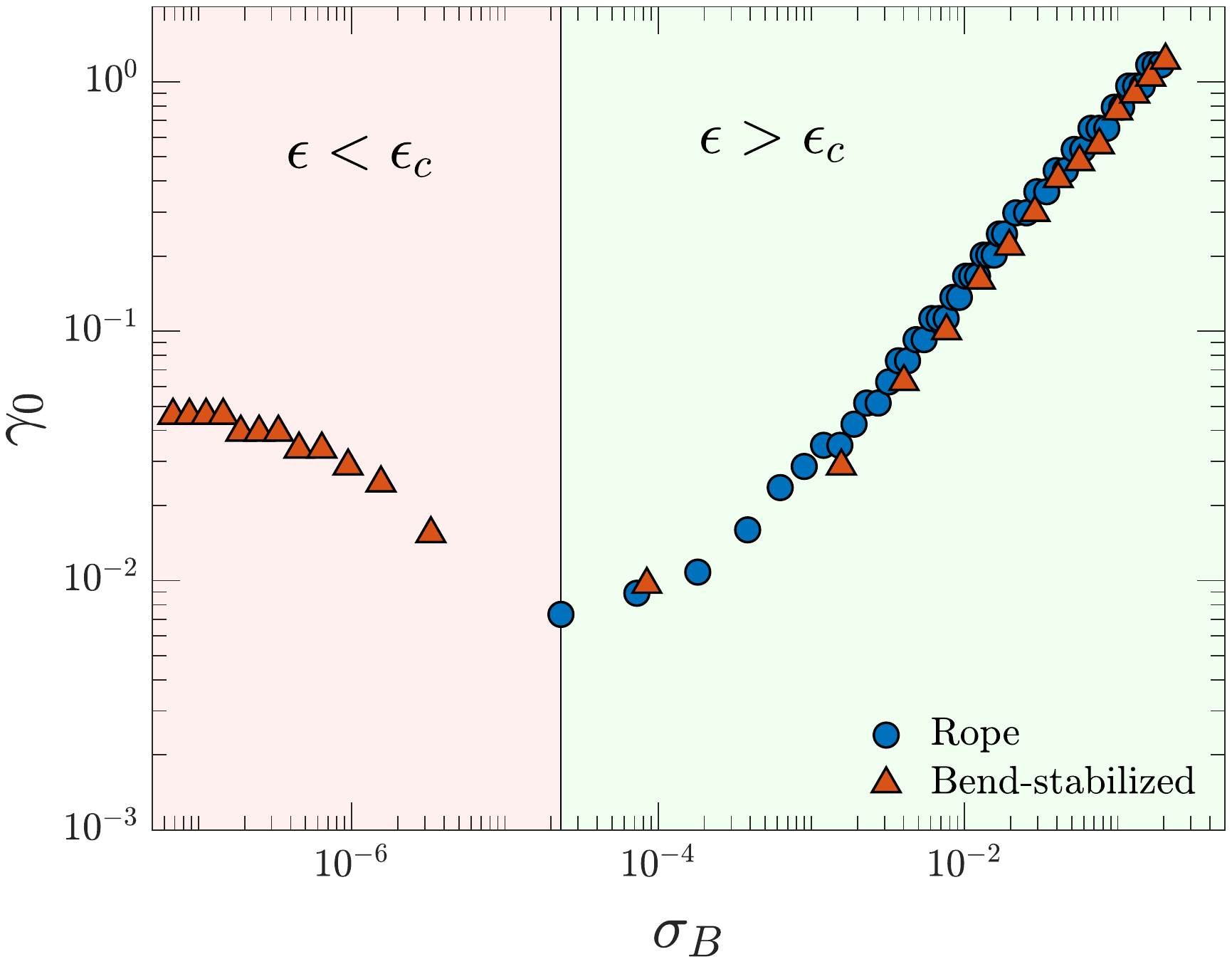}
	\caption{ \label{fig:5} The effect of prestress on the onset of strain stiffening. The data is obtained using the same network as in figure \ref{fig:4} and prestress is applied by bulk expansion. Onset of strain stiffening $\gamma_0$ which is defined as the strain where $K \simeq 2G$ increases by applying external stress to a rope network. It is noted that rope networks are unstable if the applied bulk strain is less than $\epsilon_c$, hence, based on our definition, $\gamma_0$ for rope networks is undefined in the regime where $\epsilon < \epsilon_c$. Bend-stabilized networks show a non-monotonic behavior, $\gamma_0$ decreases for small extensional strains $\epsilon<\epsilon_c$ in which the network is still bending-dominated and increases after applying $\epsilon> \epsilon_c$ in which the network is stretching-dominated.}
\end{figure}

Interestingly, the onset of strain stiffening $\gamma_0$ which we define the strain where differential shear modulus is twice as large the linear shear modulus $K \simeq 2 G$, increases with increasing external normal stresses for a sub-isostatic network in a rope-like potential (see Fig. \ref{fig:5}). In the case of bend-stabilized networks, however, $\gamma_0$ decreases by applying extensional strain $\epsilon<\epsilon_c$ and shows a similar behavior to rope networks under extensional strains larger than $\epsilon_c$. In the experimental studies of biopolymer networks, one could apply external normal stress $\sigma_P$ and measure the differential shear modulus $K$ at this stressed state. By measuring shear modulus curves for every applied prestress, onset of strain stiffening $\gamma_0$ versus the prestress $\sigma_P$ can be obtained and could serve as an indicator that such networks have been strained past the critical strain $\epsilon_c$. So dependence of $\gamma_0$ on $\sigma_P$ can serve as an indicator of whether the network is bending-dominated or stretching-dominated in the reference state. Indeed the experimental studies on the reconstituted collagen networks are shown to develop normal stresses due to polymerization and boundary effects \cite{licup_stress_2015} which can significantly affect their mechanical response. In real tissues, the external stresses exist due to the interactions between different tissues as well as embedded cells.

\section{Conclusion}
Floppy sub-isostatic fiber networks are stabilized via various mechanisms such as applying large strain, introducing fiber bending interactions, imposing active stresses, and thermal fluctuations. Here we have shown that external normal stresses can rigidify linearly unstable fiber networks. We considered the case in which connected network nodes interact only through rope-like tensile forces. The stability of rope-like structures has been studied previously in the context of disordered networks \cite{delaney_onset_2005,shen_statistical_2006,wang_interplay_2011,han_cell_2018,ronceray_stress-dependent_2019}, although the role of prestress on the shear rheology of such systems was not examined. We found that the linear shear modulus of these stress-stabilized networks scales as a power law with the applied external normal stress. The scaling exponent exhibits a non-mean-field value for connectively disordered networks and  a mean field value for connectively homogeneous structures. We also investigated the effect of prestress on the criticality of the networks. In order to stabilize a sub-isostatic rope network,  a non-zero prestress which corresponds to an extension strain larger than $\epsilon_c$ needs to be applied. This indicates that the network becomes rigid and the criticality corresponding to the transition between two floppy and rigid states is removed. Indeed, the non-affine fluctuations of a stress-stabilized rope network show no peak and are clearly supressed. For a bend-stabilized network, however, small prestress corresponding to a small extension $\epsilon < \epsilon_c$ shifts the critical strain $\gamma_c$ to lower values which is clearly observed by calculating the non-affine fluctuations. Moreover, we find that the onset of strain stiffening $\gamma_0$ monotonically increases by applying prestress to a rope network. For a bend-stabilized network, however, this behavior is non-monotonic with a decreasing trend in the bending-dominated regime and increasing trend similar to the stress-stabilized rope networks in the stretching-dominated regime. The distinctive behavior of $\gamma_0$ versus external stress $\sigma_P$ can be used to determine whether a fiber network is in the bending-dominated or stretching-dominated regime.

\section*{Acknowledgments}

This work was supported in part by the National Science Foundation Division of Materials Research (Grant DMR1826623) and the National Science Foundation Center for Theoretical Biological Physics (Grant PHY-1427654).

%\bibliography{RopeCitations}


\end{document}